\documentclass[final,3p,times,numbers,sort&compress]{elsarticle}

\usepackage{amsmath}
\usepackage{amsfonts}
\usepackage{amsmath}
\usepackage{amssymb}
\usepackage{amsthm}
\usepackage[utf8]{inputenc}
\usepackage{graphicx}
\usepackage{bm}
\usepackage{bbm}

\journal{Physica A}

\usepackage[breaklinks=true,colorlinks=true,linkcolor=blue,urlcolor=blue,citecolor=blue]{hyperref}

\newcommand{\defeq}{\mathrel{\mathop:}=}
\newcommand{\U}{\mathcal{U}}
\renewcommand{\H}{\mathcal{H}}
\newcommand{\params}{\boldsymbol{\lambda}}

\begin{document}

\begin{frontmatter}

\title{Kappa distributions as asymptotic marginals of exponential family ensembles}

\author[cchen,unab]{Sergio Davis\corref{cor1}}
\address[cchen]{Research Center on the Intersection of Plasma Physics, Matter and Complexity (P$^2$mc),\\ Comisión Chilena de Energía Nuclear, Casilla 188-D, Santiago, Chile}
\address[unab]{Departamento de Física y Astronomía, Facultad de Ciencias Exactas, Universidad Andres Bello,\\ Sazié 2212, piso 7, 8370136, Santiago, Chile.}
\ead{sergio.davis@cchen.cl}

\cortext[cor1]{Corresponding author}

\begin{abstract}
Recently (Physica A \textbf{660} 130370, 2025), emergence of superstatistical behavior in driven classical systems has been shown for systems with constant microcanonical heat capacity. 
As an application of this result, in this work we show that a system of $N$ particles with inverse gamma distribution of total kinetic energies must have kappa-distributed single-particle 
velocities in the limit $N \rightarrow \infty$. Our results provide insight into the nature of kappa distributions outside the theory of nonextensive statistical mechanics, while also
bringing forward a practical method for the generation of kappa velocities via Monte Carlo Metropolis simulation in the inverse gamma ensemble.
\end{abstract}

\end{frontmatter}

\section{Introduction}

The kappa distribution~\cite{Pierrard2010, Livadiotis2017, Lazar2021} describes the single-particle velocities in collisionless, non-equilibrium plasmas as found, for instance, in the magnetosphere of 
the Earth~\cite{Antonova2008, Espinoza2018, Kirpichev2020, Eyelade2021} and other planets~\cite{Carbary2014,Nicolaou2020}, in the solar wind~\cite{Maksimovic1997b, Nicolaou2019, ZentenoQuinteros2021} and 
other types of space plasmas~\cite{Raymond2010, Nicholls2017}. It can be written as
\begin{equation}
\label{eq:kappa}
P(\bm v|\kappa, \theta) = \left[\pi\left(\kappa-\frac{3}{2}\right) \theta^2\right]^{-\frac{3}{2}} \frac{\Gamma(\kappa+1)}{\Gamma\left(\kappa-\frac{1}{2}\right)}
\left[1 + \frac{1}{\kappa-\frac{3}{2}}\left(\frac{\bm{v}^2}{\theta^2}\right)\right]^{-(\kappa+1)},
\end{equation}
where $\theta$ is the \emph{thermal velocity} and $\kappa$ is known as the spectral index. In the limit $\kappa \rightarrow \infty$, the kappa distribution in \eqref{eq:kappa} reduces to the Maxwellian 
distribution
\begin{equation}
P(\bm v|\beta) = \left(\frac{m\beta}{2\pi}\right)^{\frac{3}{2}}\exp\left(-\frac{\beta m\bm{v}^2}{2}\right),
\end{equation}
with inverse temperature
\begin{equation}
\beta \defeq \frac{1}{k_B T} = \frac{2}{m\theta^2}.
\end{equation}

Recently~\cite{Davis2023e, Davis2026}, we have introduced a different, but completely equivalent, parameterization of the kappa distribution, namely
\begin{equation}
\label{eq:kappa_new}
P(\bm v|u, \beta_S) = \frac{\Gamma\Big(\frac{3}{2}+\frac{1}{u}\Big)}{\Gamma\Big(\frac{1}{u}\Big)}
\,\left(\frac{m u\beta_S}{2\pi}\right)^{\frac{3}{2}}\left[1 + u\beta_S\left(\frac{m\bm{v}^2}{2}\right)\right]^{-\left(\frac{1}{u}+\frac{3}{2}\right)}.
\end{equation}
where the new parameter $u$ is related to the spectral index $\kappa$ by 
\begin{equation}
\kappa = \frac{1}{u}+\frac{1}{2},
\end{equation}
while $\beta_S$ is related to $\theta$ by
\begin{equation}
\theta^2 = \frac{2}{m \beta_S(1- u)}.
\end{equation}

One advantage of this parameterization comes from the theory of superstatistics~\cite{Beck2003, Beck2004}. In this framework, \eqref{eq:kappa_new} is obtained as a continuous mixture of canonical distributions, i.e.
\begin{equation}
P(\bm v|u, \beta_S) = \int_0^\infty d\beta\,P(\beta|u, \beta_S)\left(\frac{m\beta}{2\pi}\right)^{\frac{3}{2}}\exp\left(-\frac{\beta m\bm{v}^2}{2}\right),
\end{equation}
where $P(\beta|u, \beta_S)$ is a gamma distribution of inverse temperatures,
\begin{equation}
P(\beta|u, \beta_S) = \frac{1}{u\beta_S\,\Gamma\left(\frac{1}{u}\right)}\exp\left(-\frac{\beta}{u\beta_S}\right)\left(\frac{\beta}{u\beta_S}\right)^{\frac{1}{u}-1}.
\end{equation}

Here, $\beta_S$ and $u$ correspond precisely to the mean and relative variance, respectively, of the superstatistical distribution of inverse temperatures, that is,
\begin{equation}
\beta_S = \big<\beta\big>_{u, \beta_S},
\end{equation}
and
\begin{equation}
u = \frac{\big<\beta^2\big>_{u, \beta_S}}{(\beta_S)^2} - 1.
\end{equation}

Several explanations of the origin of kappa distributions have been proposed, including but not limited to Tsallis' nonextensive statistical mechanics~\cite{Tsallis1988, Tsallis2009c} and superstatistics
~\cite{Beck2003, Beck2004}, the latter recently gaining traction in the plasma physics community~\cite{Ourabah2015, Ourabah2020b, Ourabah2024, Davis2026}. However, none of them seem to be the definitive answer. 
In this work, based on the recent proof~\cite{Davis2025} connecting superstatistical behavior with the thermodynamic limit of driven classical systems, we present another piece of the picture, by showing analytically 
and numerically how the kappa distribution arises naturally from an inverse-gamma distribution of total kinetic energy for a system of $N$ particles with large $N$. The validity of this fact does not rely on 
superstatistics, the principle of maximum entropy or information-theoretical concepts, as it is simply a consequence of marginalization in the space of velocities. No distribution of temperatures or definitions 
of entropy are required. Metropolis Monte Carlo simulations, as well as direct numerical computations, show that in practice, ``large $N$'' can be achieved with as few as 50 particles.

This paper is organized as follows. Section \ref{sec:ess} presents the definitions of energy steady states, microcanonical and fundamental inverse temperatures, as well as the invariants $u$ and $\beta_S$.
Then, in Section \ref{sec:invgamma} we present the inverse gamma ensemble for a system of $N$ particle velocities, while Section \ref{sec:marginal} computes the exact single-particle velocity distribution in 
the inverse gamma ensemble, showing that it reduces to the kappa distribution in the thermodynamic limit. Section \ref{sec:metropolis} validates the previous results against Metropolis Monte Carlo simulation, 
and we finish with some concluding remarks in Section \ref{sec:concluding}.

\section{Inverse temperature invariants in energy steady states including superstatistics}
\label{sec:ess}

\noindent
We will consider steady states as the time-independent solutions of the Liouville equation,
\begin{equation}
\frac{\partial \mathcal{P}_t(\bm \Gamma)}{\partial t} + \Big\{\mathcal{P}_t(\bm \Gamma), \H(\bm \Gamma)\Big\} = 0,
\end{equation}
where $\mathcal{P}_t(\bm \Gamma) \defeq P(\bm{\Gamma}_t = \bm \Gamma|\params)$ is the probability density of the system microstate being $\bm \Gamma$ at the time $t$. This means that steady states are 
solutions of
\begin{equation}
\label{eq:steady}
\Big\{\mathcal{P}(\bm \Gamma), \H(\bm \Gamma)\Big\} = 0.
\end{equation}

\noindent
In particular, we will define the \emph{energy steady states} as solutions of \eqref{eq:steady} of the form
\begin{equation}
\mathcal{P}(\bm \Gamma) = \rho\big(\H(\bm \Gamma); \params\big),
\end{equation}
where $\params$ is the particular set of parameters that describes the steady state. That is, in energy steady states, the energy of the microstate defines its probability density.
For all energy steady states, we can define two inverse temperature estimators. On the one hand, the \emph{fundamental inverse temperature}
\begin{equation}
\beta_F(E; \params) \defeq -\frac{\partial}{\partial E}\ln \rho(E; \params)
\end{equation}
depends explicitly on the ensemble function, while the \emph{microcanonical inverse temperature}
\begin{equation}
\beta_\Omega(E) = \frac{\partial}{\partial E}\ln \Omega(E)
\end{equation}
where $\Omega(E) \defeq \int d\bm{\Gamma}\,\delta\big(\H(\bm \Gamma)-E\big)$ is the density of states, is an intrinsic estimator that only depends on the shape of the Hamiltonian.
Moreover, in every energy steady state it holds that 
\begin{equation}
\beta_S \defeq \big<\beta_F\big>_{\params} = \big<\beta_\Omega\big>_{\params}.
\end{equation}

\noindent
The \emph{inverse temperature covariance}~\cite{Davis2022b} $\U$ is defined as
\begin{equation}
\U \defeq \big<\delta\beta_F\,\delta\beta_\Omega\big>_{\params}
\end{equation}
but it can also be written as
\begin{equation}
\U = \big<(\delta \beta_\Omega)^2\big>_{\params} + \big<{\beta_\Omega}'\big>_{\params} = \big<(\delta \beta_F)^2\big>_{\params} - \big<{\beta_\Omega}'\big>_{\params}.
\end{equation}

\noindent
Both $\beta_S$ and $\U$ are in fact properties of the inverse temperature distribution in superstatistics. In that case,
\begin{align}
\beta_S & = \big<\beta\big>_{\params}, \\
\U & = \big<(\delta \beta)^2\big>_{\params}
\end{align}
for every superstatistical model. In other words, the mean of the superstatistical $\beta$ coincides with the mean of both $\beta_\Omega$ and $\beta_F$, and $\U$ coincides with the variance of the 
superstatistical $\beta$. From $\U$ and $\beta_S$ we can define the reduced inverse temperature covariance
\begin{equation}
u \defeq \frac{\U}{(\beta_S)^2},
\end{equation}
which is a dimensionless quantity. As a covariance, $\U$ can be negative for energy steady states outside of superstatistics. We denote those states as \emph{subcanonical}, whereas states with $\U > 0$ 
(including superstatistical states) are \emph{supercanonical}.

\section{The inverse gamma kinetic ensemble}
\label{sec:invgamma}

\noindent
We will consider a system of $N$ particles in three dimensions, and will denote the $3N$-dimensional vector of all velocity components as $\bm{V} \defeq (\bm{v}_1, \ldots, \bm{v}_N)$. 
Furthermore, we will assume an energy steady state where the probability density for $\bm V$ is given by
\begin{equation}
\label{eq:invgamma_ens}
P(\bm V|\gamma, K_0) = \frac{1}{\eta(\gamma, K_0)}\exp\left(-\frac{K_0}{K(\bm V)}\right)\left(\frac{K_0}{K(\bm V)}\right)^{\gamma+\frac{3N}{2}}
\end{equation}
with parameters $\gamma > 0$ and $K_0 > 0$, and where $K(\bm V)$ the total kinetic energy
\begin{equation}
\label{eq:kin}
K(\bm V) \defeq \sum_{i=1}^N \frac{m\bm{v}_i^2}{2}.
\end{equation}

\noindent
The corresponding probability density for $K$ is computed as
\begin{equation}
P(K|\gamma, K_0) = \int d\bm{V}\,P(\bm V|\gamma, K_0)\,\delta\left(\sum_{i=1}^N \frac{m\bm{v}_i^2}{2} - K\right),
\end{equation}
which, by using the density of states of $K(\bm V)$ in \eqref{eq:kin},
\begin{equation}
\label{eq:omega_kin}
\Omega_K(K; N) \defeq \int d\bm{V}\,\delta\left(\sum_{i=1}^N \frac{m\bm{v}_i^2}{2} - K\right) = W_N\,K^{\frac{3N}{2}-1}
\end{equation}
where
\begin{equation}
\label{eq:wn}
W_N \defeq \frac{1}{\Gamma\left(\frac{3N}{2}\right)} \left(\frac{2\pi}{m}\right)^{\frac{3N}{2}}
\end{equation}
and imposing normalization, reduces to
\begin{equation}
\label{eq:invgammakin}
P(K|\gamma, K_0) = \frac{1}{K_0\,\Gamma(\gamma)}\exp\left(-\frac{K_0}{K}\right)\,\left(\frac{K_0}{K}\right)^{\gamma+1},
\end{equation}
fixing the value of $\eta(\gamma, K_0)$ to be
\begin{equation}
\eta(\gamma, K_0) = W_N\,\Gamma(\gamma)(K_0)^{\frac{3N}{2}}
\end{equation}

\noindent
The moments of this distribution are given by
\begin{equation}
\label{eq:moments}
\big<K^n\big>_{\gamma, K_0} = \prod_{m=1}^n \left(\frac{K_0}{\gamma-m}\right) = (K_0)^n \;\frac{\Gamma(\gamma-n)}{\Gamma(\gamma)},
\end{equation}
for $n < \gamma$, as shown in \ref{sec:appendix}. In particular, $n = 1$ gives the mean kinetic energy
\begin{equation}
\big<K\big>_{\gamma, K_0} = \frac{K_0}{\gamma-1}
\end{equation}
while $n = 2$ yields
\begin{equation}
\big<K^2\big>_{\gamma, K_0} = \frac{(K_0)^2}{(\gamma-1)(\gamma-2)},
\end{equation}
therefore the relative variance of $K$ is
\begin{equation}
\frac{ \big<(\delta K)^2\big>_{\gamma, K_0} }{ \big<K\big>_{\gamma, K_0}^2 } = \frac{1}{\gamma-2}.
\end{equation}

Now let us compute the invariant parameters $u$ and $\beta_S$ for the ensemble in \eqref{eq:invgamma_ens}. We obtain the microcanonical inverse temperature by taking the logarithmic derivative of 
\eqref{eq:omega_kin},
\begin{equation}
\label{eq:betaom}
\beta_\Omega(K) = \frac{3N-2}{2K}
\end{equation}
so we can compute $\beta_S$ simply as
\begin{equation}
\label{eq:betaS_exact}
\beta_S = \big<\beta_\Omega\big>_{\gamma, K_0} = \frac{3N-2}{2}\big<K^{-1}\big>_{\gamma, K_0} = \frac{(3N-2)\gamma}{2 K_0},
\end{equation}
where we have used $n = -1$ in \eqref{eq:moments} to obtain
\begin{equation}
\big<K^{-1}\big>_{\gamma, K_0} = \frac{\gamma}{K_0}.
\end{equation}

\noindent
Similarly, using $n = -2$ we obtain
\begin{equation}
\big<K^{-2}\big>_{\gamma, K_0} = \frac{\gamma(\gamma+1)}{(K_0)^2}
\end{equation}
from which we readily compute
\begin{equation}
\big<(\delta \beta_\Omega)^2\big>_{\gamma, K_0} = \left<\left(\frac{3N-2}{2K}\right)^2\right>_{\gamma, K_0} - (\beta_S)^2 = \frac{(3N-2)^2 \gamma}{4(K_0)^2}
\end{equation}
and
\begin{equation}
\big<{\beta_\Omega}'\big>_{\gamma, K_0} = \left<\frac{2-3N}{2K^2}\right>_{\gamma, K_0} = \frac{(2-3N)\gamma(\gamma+1)}{2(K_0)^2},
\end{equation}
finally arriving at
\begin{equation}
\label{eq:U}
\U = \frac{\gamma(3N-2)\big(3N-2(\gamma+2)\big)}{4(K_0)^2}.
\end{equation}

\noindent
In summary, the mean inverse temperature $\beta_S$ is
\begin{equation}
\beta_S = \frac{(3N-2)\gamma}{2K_0}
\end{equation}
while the reduced inverse temperature covariance $u$ is given by
\begin{equation}
\label{eq:u_exact}
u = \frac{3N-2\gamma-4}{\gamma(3N-2)}.
\end{equation}

\noindent
Note that $u$ can be negative, for small $N$ or large $\gamma$, in fact whenever
\begin{equation}
\frac{3N}{2} < \gamma + 2,
\end{equation}
and, in this case, the steady state is subcanonical, thus outright incompatible with superstatistics. It is also clear that, for $N \rightarrow \infty$, we have
\begin{subequations}
\label{eq:ubetaS_lim}
\begin{align}
\lim_{N \rightarrow \infty} u(\gamma, N) & = \frac{1}{\gamma}, \\
\lim_{N \rightarrow \infty} \beta_S(\gamma, K_0, N) & = \frac{3\gamma}{2 k_0},
\end{align}
\end{subequations}
with 
\begin{equation}
k_0 \defeq \lim_{N \rightarrow \infty} \frac{K_0}{N},
\end{equation}
thus, in this limit, $u > 0$. On the other hand, we can solve for $\gamma$ as a function of $u$ and $N$, obtaining
\begin{equation}
\gamma = \frac{3N-4}{(3N-2)u + 2},
\end{equation}
thus we see that
\begin{equation}
\frac{ \big<(\delta K)^2\big>_{\gamma, K_0} }{ \big<K\big>_{\gamma, K_0}^2 } = \frac{(3N-2)u + 2}{3N(1-2u) + 4u - 8},
\end{equation}
which for large $N$ reduces to
\begin{equation}
\lim_{N \rightarrow \infty} \frac{ \big<(\delta K)^2\big>_{\gamma, K_0} }{ \big<K\big>_{\gamma, K_0}^2 } = \frac{u}{1-2u},
\end{equation}
implying that, in that limit,
\begin{equation}
0 < u < \frac{1}{2}.
\end{equation}

We can check the internal consistency of our calculations by using the fundamental inverse temperature $\beta_F$ of the ensemble in \eqref{eq:invgamma_ens},
\begin{equation}
\beta_F(K; \gamma, K_0) = \frac{3N + 2\gamma}{2K} - \frac{K_0}{K^2}.
\end{equation}

\noindent
By combining it with \eqref{eq:betaom}, we have that
\begin{equation}
\big<\beta_F\,\beta_\Omega\big>_{\gamma, K_0} = \frac{3N-2}{2}\left[\frac{(3N+2\gamma)}{2}\big<K^{-2}\big>_{\gamma, K_0} - K_0\big<K^{-3}\big>_{\gamma, K_0}\right]
= \frac{3N-2}{2}\cdot \frac{(3N-4)\gamma(\gamma+1)}{2(K_0)^2},
\end{equation}
hence the covariance between $\beta_F$ and $\beta_\Omega$ is
\begin{equation}
\big<\delta\beta_F\,\delta\beta_\Omega\big>_{\gamma, K_0} = \big<\beta_F\,\beta_\Omega\big>_{\gamma, K_0} - (\beta_S)^2 = \frac{\gamma(3N-2)\big(3N-2(\gamma+2)\big)}{4(K_0)^2},
\end{equation}
exactly equal to $\U$ in \eqref{eq:U}. 

\section{Computation of the single-particle velocity distribution}
\label{sec:marginal}

\noindent
Now let us write $\bm{V} = (\bm{v}, \bm{\tilde{V}})$, where $\bm{v} \defeq \bm{v}_1$ and $\bm{\tilde{V}} \defeq (\bm{v}_2, \bm{v}_3, \ldots, \bm{v}_N)$. With this choice of vectors we can write our original 
$N$-particle distribution in \eqref{eq:invgamma_ens} as the joint distribution of $\bm v$ and $\bm{\tilde{V}}$,
\begin{equation}
P(\bm{v}, \bm{\tilde{V}}|\gamma, K_0) = \frac{(K_0)^{-\frac{3N}{2}}}{W_N\,\Gamma(\gamma)}\exp\left(-\frac{K_0}{k(\bm v) + \tilde{K}(\bm{\tilde{V}})}\right)\left(\frac{K_0}{k(\bm v) + \tilde{K}(\bm{\tilde{V}})}\right)^{\gamma+\frac{3N}{2}},
\end{equation}
with 
\begin{equation}
\tilde{K}(\bm{\tilde{V}}) \defeq \sum_{i=2}^N \frac{m\bm{v}_i^2}{2},
\end{equation}
and furthermore, we can obtain the distribution of $\bm{v}$ by the marginalization rule,
\begin{equation}
P(\bm{v}|\gamma, K_0) = \int d\bm{\tilde{V}}\,P(\bm{v}, \bm{\tilde{V}}|\gamma, K_0).
\end{equation}

By replacing the integration over $\bm{\tilde{V}}$ by integration over the kinetic energy $\tilde{K}$ of the $N$-1 particles weighted by the density of states $\Omega_K(\tilde{K}; N-1)$, we have
\begin{equation}
\begin{split}
P(\bm{v}|\gamma, K_0) & = \frac{(K_0)^{-\frac{3N}{2}}}{W_N\,\Gamma(\gamma)}\int_0^\infty d\tilde{K}\,\Omega_K(\tilde{K}; N-1)\exp\left(-\frac{K_0}{k(\bm v) + \tilde{K}}\right)\left(\frac{K_0}{k(\bm v) + \tilde{K}}\right)^{\gamma+\frac{3N}{2}} \\
& = \frac{W_{N-1}\,(K_0)^{-\frac{3N}{2}}}{W_N\,\Gamma(\gamma)}\int_0^\infty d\tilde{K}\,\tilde{K}^{\frac{3N}{2}-\frac{5}{2}}\,\exp\left(-\frac{K_0}{k(\bm v) + \tilde{K}}\right)\left(\frac{K_0}{k(\bm v) + \tilde{K}}\right)^{\gamma+\frac{3N}{2}},
\end{split}
\end{equation}
which, upon the change of variables
\begin{displaymath}
\tilde{K} \rightarrow z \defeq \frac{k(\bm v) + \tilde{K}}{K_0}
\end{displaymath}
reads
\begin{equation}
P(\bm{v}|\gamma, K_0) = \frac{W_{N-1}}{W_N\,\Gamma(\gamma)}(K_0)^{-\frac{3}{2}}\int_0^\infty dz\,\left[z-\frac{k(\bm v)}{K_0}\right]_+^{\frac{3N}{2}-\frac{5}{2}}\,\exp\left(-\frac{1}{z}\right)
z^{-\gamma-\frac{3N}{2}}.
\end{equation}

\noindent
Using the integral
\begin{equation}
\int_c^\infty dz\,z^{-a}\,(z-c)^b\,\exp\left(-\frac{1}{z}\right) = c^{b-a+1} \frac{\Gamma(a-b-1)\Gamma(b+1)}{\Gamma(a)}\phantom{.}_1F_1\left(a-b-1, a, -\frac{1}{c}\right)
\end{equation}
where $\phantom{.}_1F_1$ is the Kummer confluent hypergeometric function~\cite{Abramowitz1972}, and replacing $W_N$ and $W_{N-1}$ according to \eqref{eq:wn}, we finally obtain
\begin{equation}
P(\bm v|\gamma, \overline{k}) = \frac{\Gamma\left(\frac{3N}{2}\right)N^{\gamma}}{\Gamma(\gamma)}\frac{\Gamma\left(\frac{3}{2}+\gamma\right)}{\Gamma\left(\frac{3N}{2}+\gamma\right)} 
\left(\frac{m}{2\pi\,k(\bm v)}\right)^{\frac{3}{2}+\gamma}\,\left(\frac{2\pi\,(\gamma-1)\overline{k}}{m}\right)^\gamma
\phantom{.}_1F_1\left(\frac{3}{2}+\gamma, \frac{3N}{2}+\gamma, -\frac{N \overline{k} (\gamma-1)}{k(\bm v)}\right).
\end{equation}
where we have defined, for convenience, the expected kinetic energy per particle $\overline{k}$ as
\begin{equation}
\overline{k} \defeq \frac{1}{N}\big<K\big>_{\gamma, K_0},
\end{equation}
such that $k_0 = \overline{k}\,(\gamma-1)$. This is the exact single-particle velocity distribution for $N \geq 2$. The corresponding single-particle kinetic energy distribution is
\begin{equation}
\label{eq:exact_kin}
P(k|\gamma, \overline{k}) = \frac{2\Gamma\left(\frac{3N}{2}\right)\Gamma\left(\frac{3}{2}+\gamma\right)}{\sqrt{\pi}\,\Gamma(\gamma)\Gamma\left(\frac{3N}{2}+\gamma\right)}\Big[N\overline{k}(\gamma-1)\Big]^{\gamma} 
k^{-(\gamma+1)}\,\phantom{.}_1F_1\left(\frac{3}{2}+\gamma, \frac{3N}{2} + \gamma, -\frac{N \overline{k}(\gamma-1)}{k}\right).
\end{equation}

\noindent
In the thermodynamic limit $N \rightarrow \infty$, we can use the limits
\begin{equation}
\lim_{N \rightarrow \infty} \frac{\Gamma\left(\frac{3N}{2}\right)N^\gamma}{\Gamma\left(\frac{3N}{2}+\gamma\right)} = \left(\frac{2}{3}\right)^\gamma
\end{equation}
and
\begin{equation}
\lim_{N \rightarrow \infty} \phantom{.}_1F_1\left(\frac{3}{2}+\gamma, \frac{3N}{2}+\gamma, -\frac{N\overline{k}(\gamma-1)}{k(\bm v)}\right) = \left[1 + \frac{2(\gamma-1)\overline{k}}{3\,k(\bm v)}\right]^{-\left(\gamma+\frac{3}{2}\right)}
\end{equation}
to obtain
\begin{equation}
\lim_{N \rightarrow \infty}\,P(\bm v|\gamma, \overline{k}) = \frac{\Gamma\left(\frac{3}{2}+\gamma\right)}{3^\gamma\,\Gamma(\gamma)}
\left(\frac{m}{2\pi\,k(\bm v)}\right)^{\frac{3}{2}+\gamma}\,\left(\frac{4\pi\,(\gamma-1)\overline{k}}{m}\right)^\gamma
\left[1 + \frac{2(\gamma-1)\overline{k}}{3\,k(\bm v)}\right]^{-\left(\gamma+\frac{3}{2}\right)},
\end{equation}
which, after some algebra, reduces to
\begin{equation}
\label{eq:kappa_pre}
\lim_{N \rightarrow \infty}\,P(\bm v|\gamma, \overline{k}) = \frac{\Gamma\left(\frac{3}{2}+\gamma\right)}{\Gamma(\gamma)}
\,\left(\frac{3 m}{4\pi\,(\gamma-1)\overline{k}}\right)^{\frac{3}{2}}\left[1 + \frac{3 m\bm{v}^2}{4(\gamma-1)\overline{k}}\right]^{-\left(\gamma+\frac{3}{2}\right)},
\end{equation}
a kappa distribution. Recalling the limit $N \rightarrow \infty$ in \eqref{eq:ubetaS_lim} we can replace
\begin{equation}
\gamma = \frac{1}{u}
\end{equation}
and
\begin{equation}
\overline{k} = \frac{3}{2(1-u)\beta_S},
\end{equation}
hence we can rewrite \eqref{eq:kappa_pre} as a function of $u$ and $\beta_S$ as
\begin{equation}
P(\bm v|u, \beta_S) \defeq \lim_{N \rightarrow \infty}\,P(\bm v|\gamma, \overline{k}) = \frac{\Gamma\left(\frac{3}{2}+\frac{1}{u}\right)}{\Gamma(\frac{1}{u})}
\,\left(\frac{m u\beta_S}{2\pi}\right)^{\frac{3}{2}}\left[1 + u\beta_S\left(\frac{m\bm{v}^2}{2}\right)\right]^{-\left(\frac{1}{u}+\frac{3}{2}\right)},
\end{equation}
which is exactly \eqref{eq:kappa_new}. The corresponding single-particle kinetic energy distribution is
\begin{equation}
\label{eq:kappa_kin}
P(k|u, \beta_S) = \frac{2(u\beta_S)^{\frac{3}{2}}\,\Gamma\left(\frac{3}{2}+\frac{1}{u}\right)}{\sqrt{\pi}\,\Gamma\left(\frac{1}{u}\right)}\Big[1 + u\beta_S k\Big]^{-\left(\frac{3}{2}+\frac{1}{u}\right)}\,\sqrt{k}.
\end{equation}

\section{Metropolis Monte Carlo simulations}
\label{sec:metropolis}

\begin{figure}[h!]
\begin{center}
\includegraphics[width=0.43\textwidth]{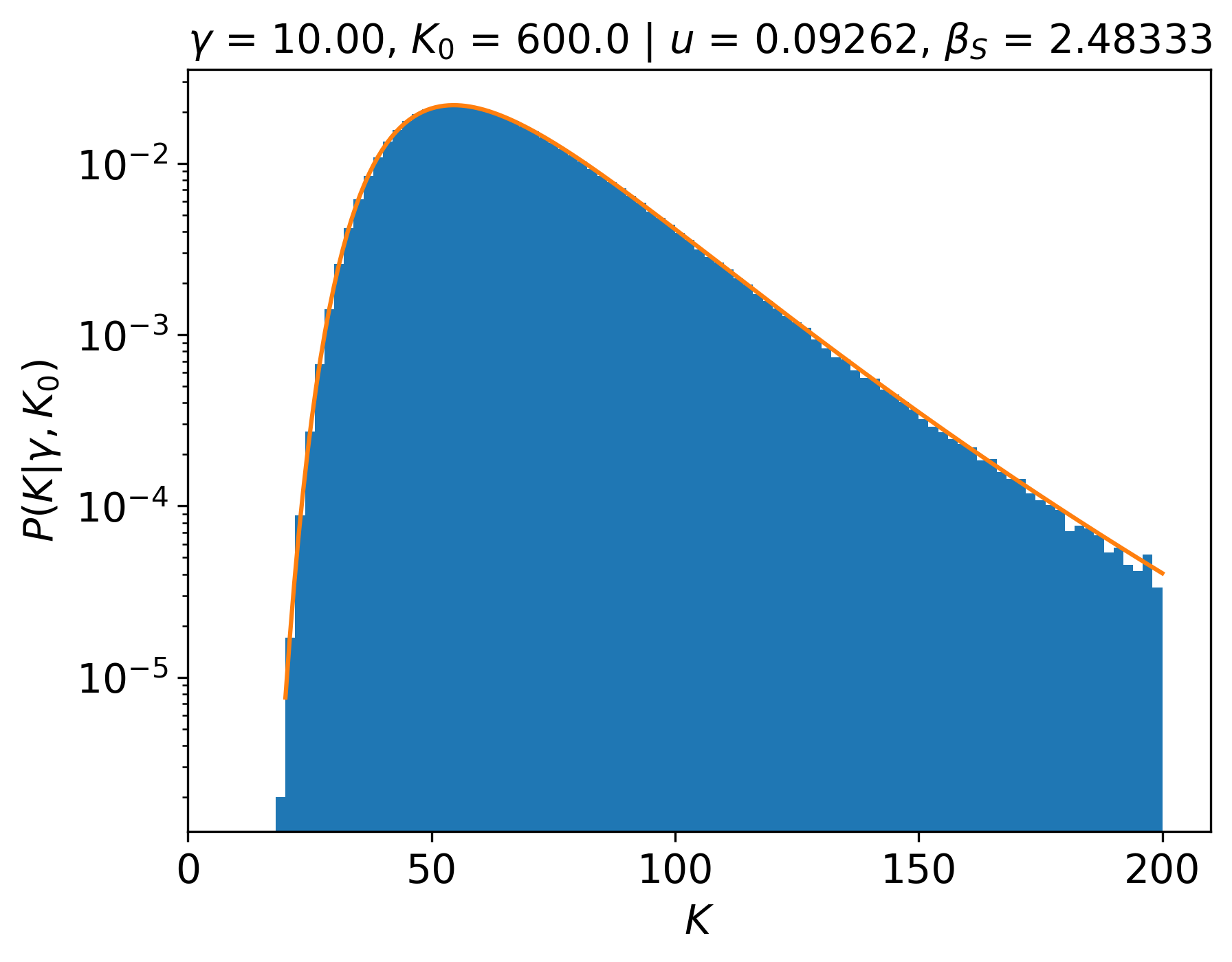}
\includegraphics[width=0.43\textwidth]{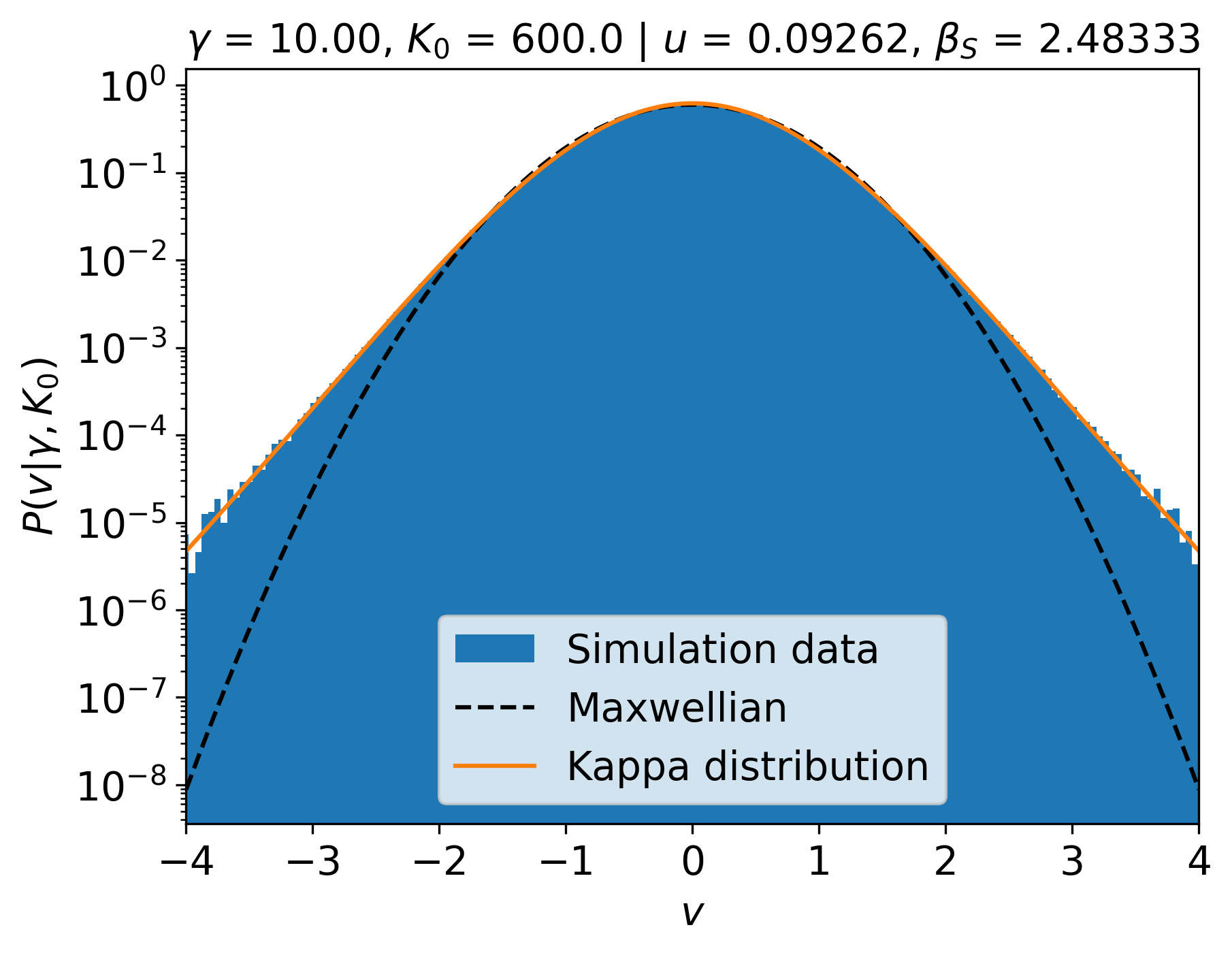}
\includegraphics[width=0.43\textwidth]{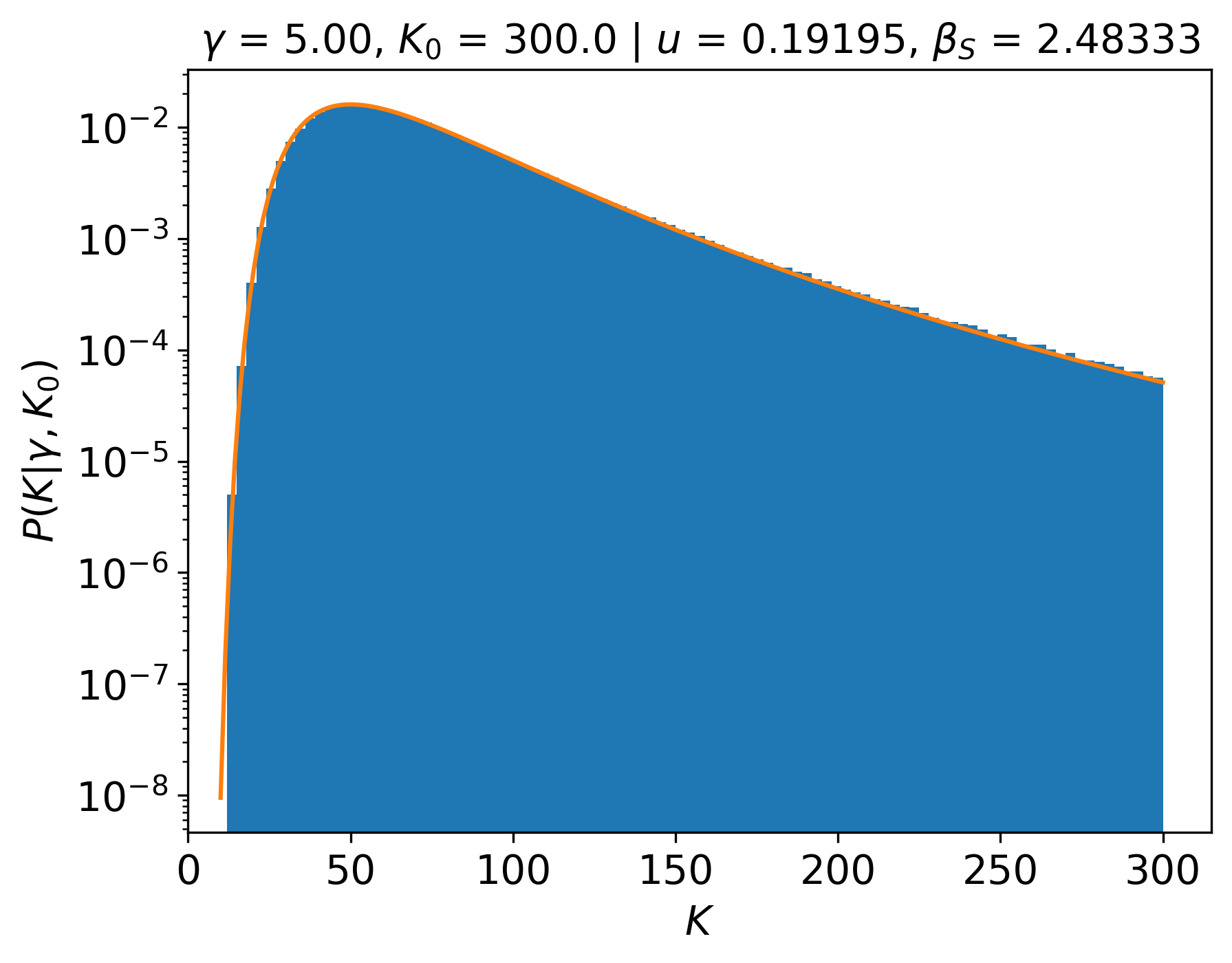}
\includegraphics[width=0.43\textwidth]{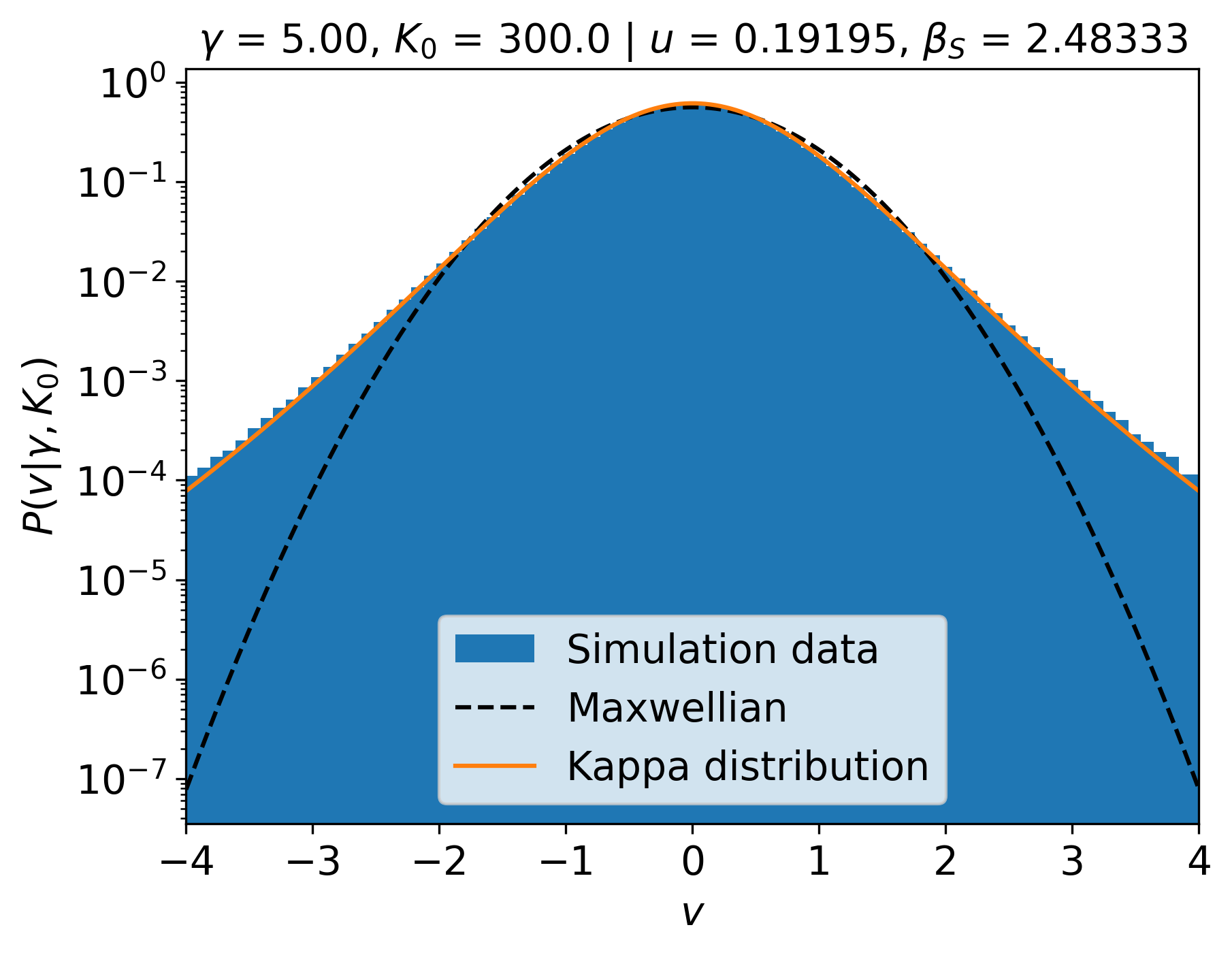}
\includegraphics[width=0.43\textwidth]{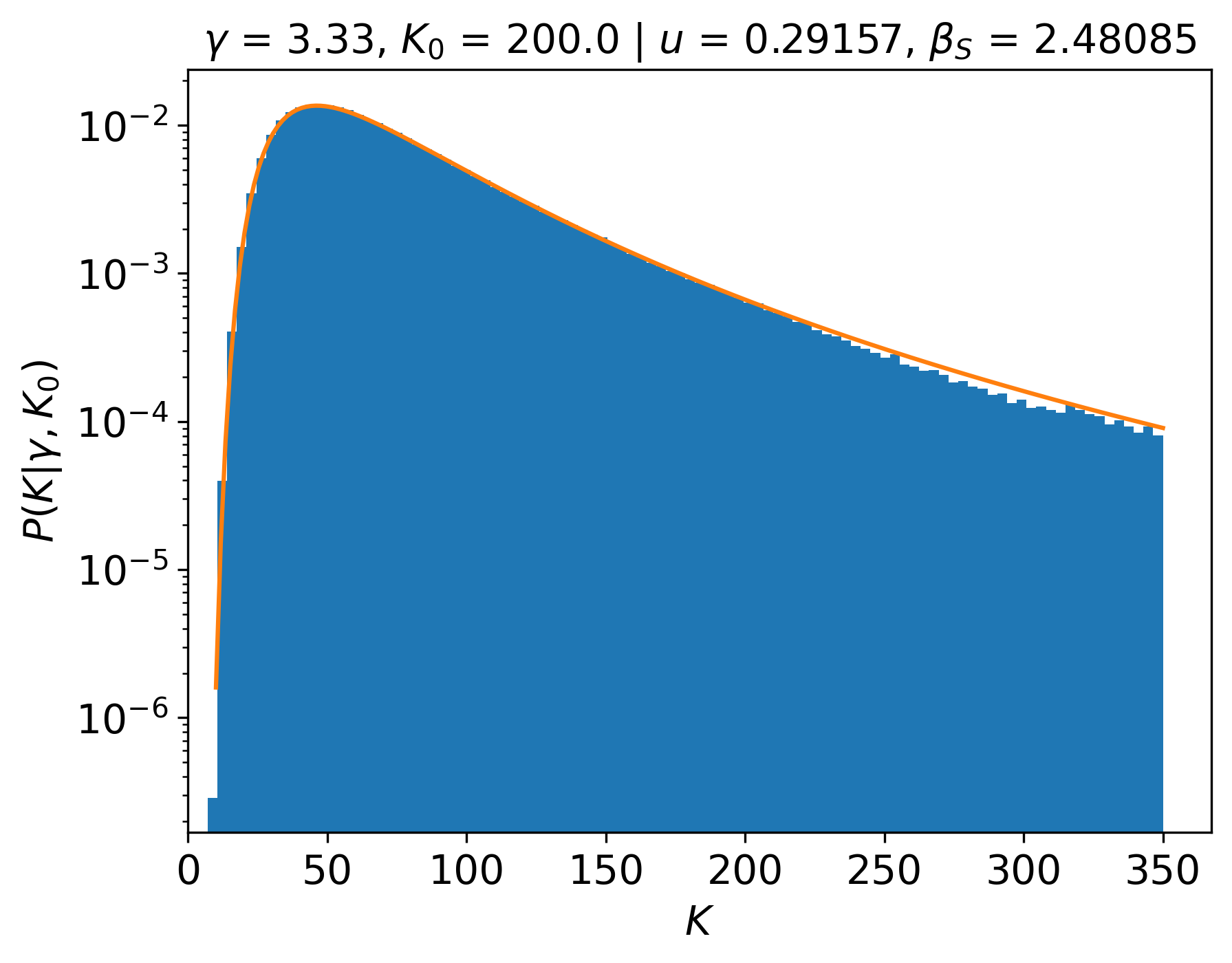}
\includegraphics[width=0.43\textwidth]{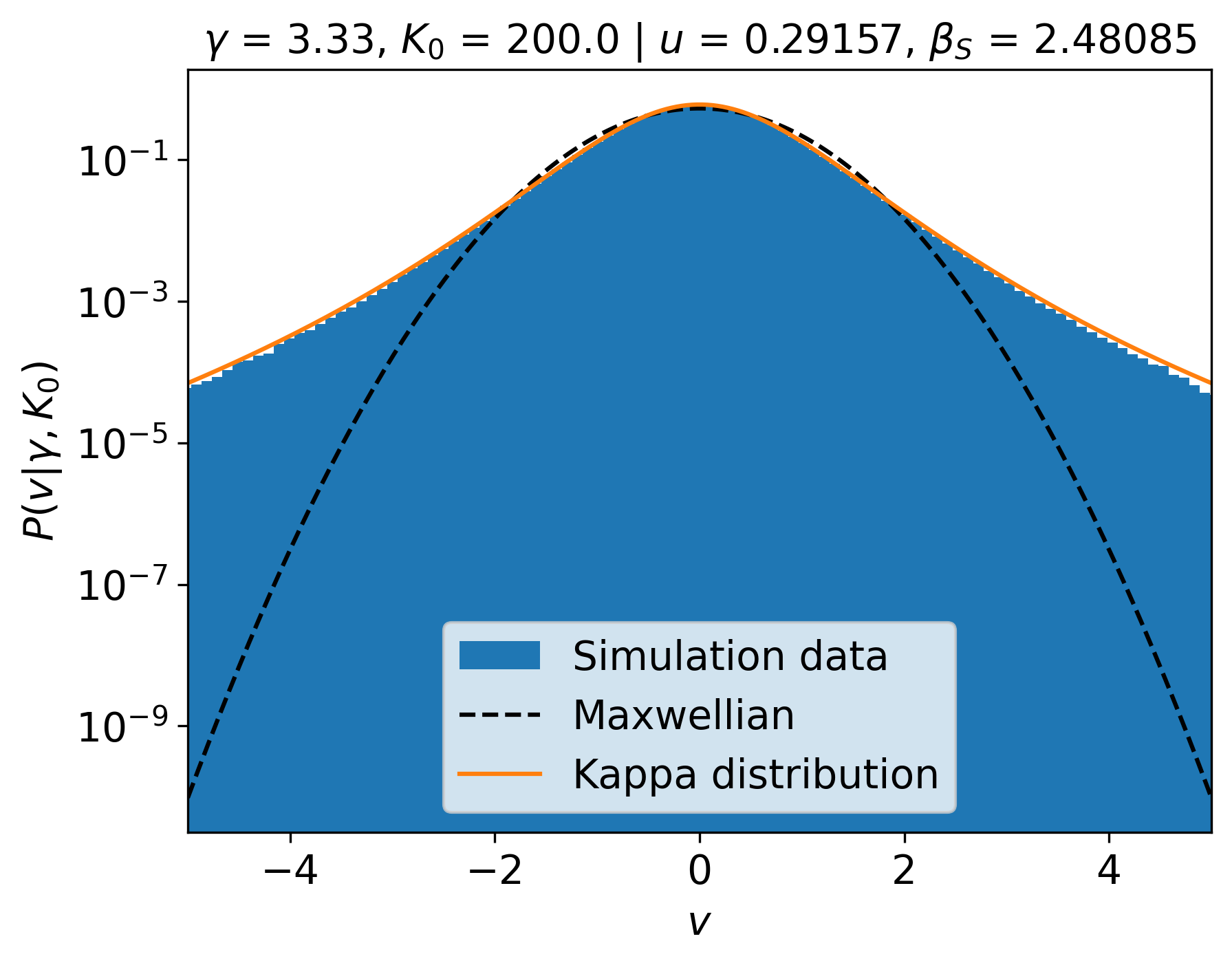}
\end{center}
\caption{Left panels, kinetic energy histograms from Metropolis simulations of a system of $N$=100 particles at $\gamma$ = 10 and $K_0$ = 600 (first row), $\gamma$ = 5 and $K_0$ = 300 (second row), and $\gamma$=3.33 
and $K_0$ = 200 (third row) together with their corresponding exact inverse-gamma distributions. Right panels, histograms of a single velocity component against one-dimensional kappa distributions for $u$ = 0.09262, $\beta_S$ = 2.48333 (first row), $u$ = 0.19195, $\beta_S$ = 2.48333 (second row), and $u$=0.29157, $\beta_S$ = 2.48085 (third row).}
\label{fig:mc}
\end{figure}

\noindent
We performed standard Metropolis simulations~\cite{Landau2015} using the acceptance probability
\begin{equation}
p_{\text{acc}} = \min\left[1, \exp\left(K_0\left[\frac{1}{K} - \frac{1}{K'}\right] + \Big(\gamma+\frac{3N}{2}\Big)\ln \frac{K}{K'}\right)\right],
\end{equation}
with $K = K(\bm V)$ and $K' = K(\bm{V}')$, where $\bm{V}'$ differs from $\bm{V}$ in the velocity of a single particle at a time. The magnitude of the random displacements on each $\bm{v}_i$ was chosen as to
keep a rejection rate between 60\% and 70\%. Figure~\ref{fig:mc} shows histograms (left panels) of kinetic energy, together with the corresponding histograms (right panels) of a single velocity component for 
$N$ = 100 particles and different values of $\gamma$ and $K_0$, corresponding roughly to $u$ = 0.1, $u$ = 0.2 and $u$ = 0.3, respectively.

\begin{figure}[t!]
\begin{center}
\includegraphics[width=0.475\textwidth]{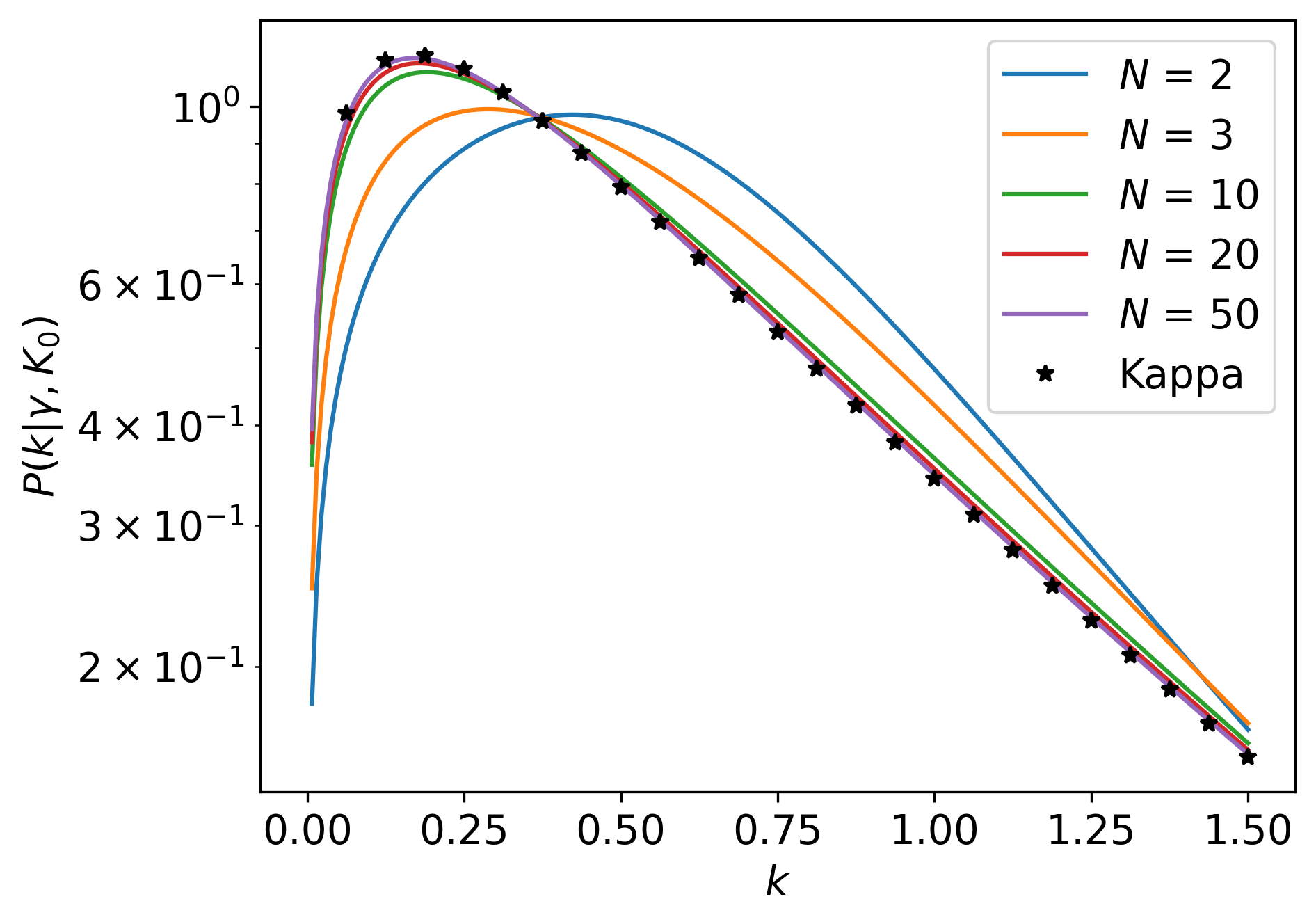}
\includegraphics[width=0.475\textwidth]{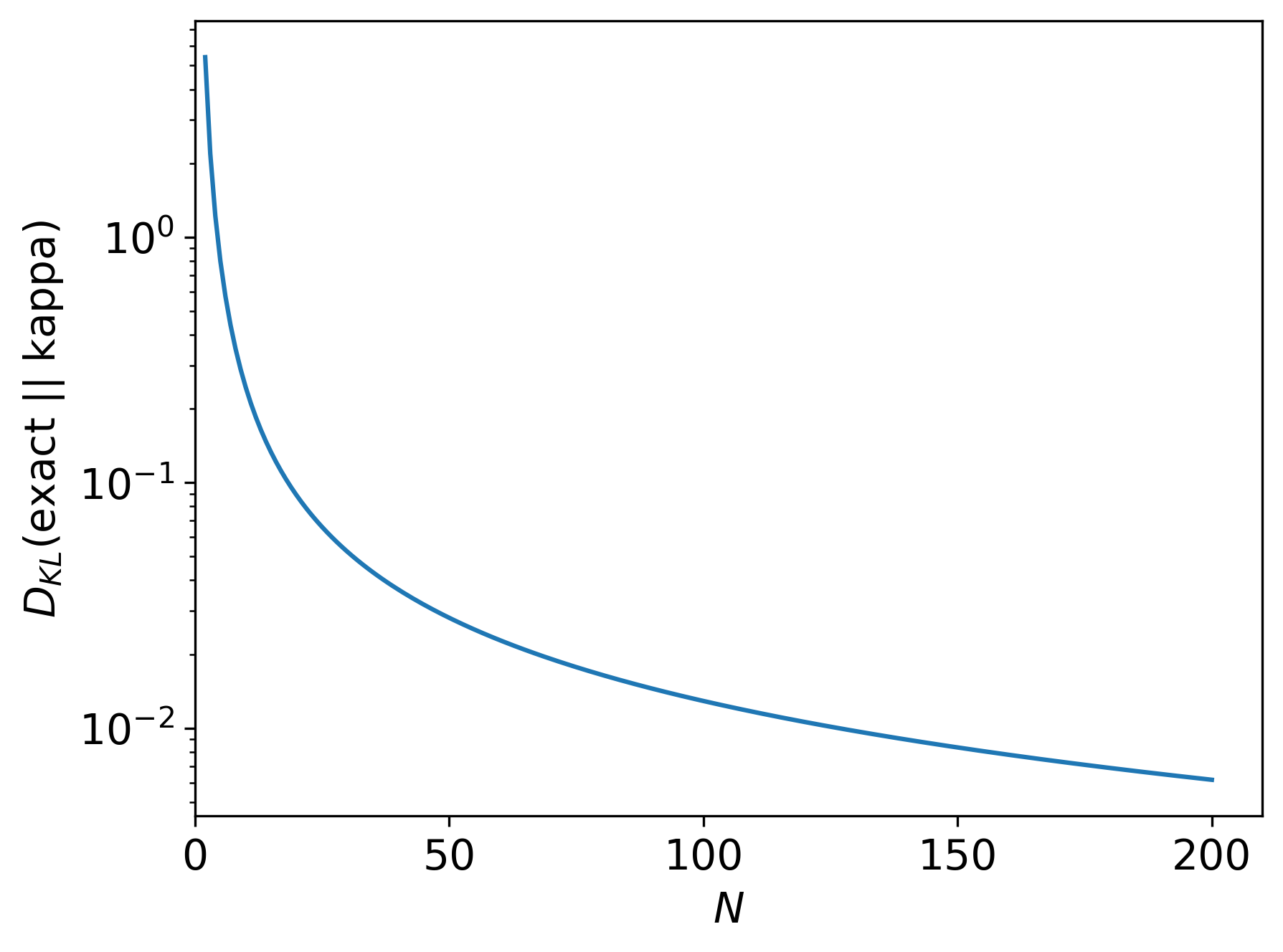}
\end{center}
\caption{Left, exact kinetic energy distribution in \eqref{eq:exact_kin} for $\gamma$ = 5, $\overline{k}$ = 0.75 and different values of $N$ (solid lines) versus the kinetic energy distribution corresponding to the 
kappa model (black stars) in \eqref{eq:kappa_kin}. Right, Kullback-Leibler divergence between the exact kinetic energy distribution in \eqref{eq:exact_kin} and the kinetic energy distribution corresponding to the kappa model in \eqref{eq:kappa_kin} as a function of $N$ for $\gamma$ = 5 and $\overline{k}$ = 0.75.}
\label{fig:ndep}
\end{figure}

\newpage
The left panel of Figure~\ref{fig:ndep} shows the convergence of the exact distribution of single-particle kinetic energies in \eqref{eq:exact_kin} towards the kappa distribution of single-particle kinetic energies in 
\eqref{eq:kappa_kin}, while the right panel shows the Kullback-Leibler distance~\cite{CoverThomas2006}
\begin{equation}
D_{KL}(\text{exact}\,||\,\text{kappa}) \defeq \int_0^\infty dk\,P(k|\gamma, \overline{k})\ln \frac{P(k|\gamma, \overline{k})}{P(k|u, \beta_S)}
\end{equation}
between those distributions as a function of $N$.

\newpage
\section{Concluding remarks}
\label{sec:concluding}

We have shown explicitly, via exact calculations and Metropolis Monte Carlo simulation, how the kappa distribution arises as the marginal single-particle distribution of an ideal gas in the inverse-gamma ensemble 
in the thermodynamic limit. In practice, 50 particles are enough to produce single-particle velocity distributions indistinguishable from kappa distributions. Our results provide a different, conceptually simple 
mechanism that generates kappa distributions without explicitly invoking a superstatistical distribution of temperatures or generalized entropies.

\section*{Acknowledgments}

\noindent
This work was supported by the NLHPC (ECM-02), Chile and FENIX (UNAB) supercomputing infrastructures.

\appendix
\section{Moments of the inverse gamma distribution}
\label{sec:appendix}

We will compute the moments of the inverse gamma distribution in \eqref{eq:invgammakin} by using the conjugate variables theorem (CVT)~\cite{Davis2012, Davis2016c},
\begin{equation}
\label{eq:cvt}
\left<\frac{\partial \omega}{\partial X}\right>_{\params} = -\left<\omega\frac{\partial}{\partial X}\ln P(X|\params)\right>_{\params}
\end{equation}
where $\omega = \omega(X)$ is a differentiable function of $X$, and the probability density $P(X|\params)$ vanishes at its boundaries. This is the case for the distribution 
in \eqref{eq:invgammakin}, whose logarithmic derivative is
\begin{equation}
\frac{\partial}{\partial K}\ln P(K|\gamma, K_0) = \frac{K_0}{K^2} - \frac{\gamma+1}{K}.
\end{equation}

\noindent
Replacing this into \eqref{eq:cvt} we have
\begin{equation}
\left<\frac{\partial \omega}{\partial K}\right>_{\gamma, K_0} = \left<\omega\left[\frac{\gamma+1}{K} - \frac{K_0}{K^2}\right]\right>_{\gamma, K_0},
\end{equation}
and with the choice
\begin{equation}
\omega(K) = K^{n+1}
\end{equation}
we readily obtain a recurrence relation
\begin{equation}
\big<K^n\big>_{\gamma, K_0} = \frac{K_0}{\gamma-n}\big<K^{n-1}\big>_{\gamma, K_0}
\end{equation}
for $n < \gamma$, with initial condition $\big<K^0\big>_{\gamma, K_0} = 1$ and solution
\begin{equation}
\big<K^n\big>_{\gamma, K_0} = \prod_{m=1}^n \left(\frac{K_0}{\gamma-m}\right) = (K_0)^n \;\frac{\Gamma(\gamma-n)}{\Gamma(\gamma)}.
\end{equation}

\bibliography{kappainvgamma}
\bibliographystyle{unsrt}

\end{document}